\documentclass[12pt]{iopart}

\usepackage{graphicx}
\usepackage{cite}
\usepackage{xcolor}

\begin{document}

\title{Bipolar electron waveguides in two-dimensional materials with tilted Dirac cones}

\author{R. R. Hartmann}
\address{
Physics Department, De La Salle University, 2401 Taft Avenue, 0922 Manila, Philippines
}
\ead{richard.hartmann@dlsu.edu.ph}

\author{M. E. Portnoi}
\address{
Physics and Astronomy, University of Exeter, Stocker Road, Exeter EX4 4QL, United Kingdom
}
\ead{m.e.portnoi@exeter.ac.uk}

\begin{abstract}
We show that the (2+1)-dimensional massless Dirac equation, which includes a tilt term, can be reduced to the biconfluent Heun equation for a broad range of scalar confining potentials, including the well-known Morse potential. Applying these solutions, we investigate a bipolar electron waveguide in 8--$Pmmn$ borophene, formed by a well and barrier, both described by the Morse potential. We demonstrate that the ability of two-dimensional materials with tilted Dirac cones to localize electrons in both a barrier and a well can be harnessed to create pseudogaps in their electronic spectrum. These pseudogaps can be tuned through varying the applied top-gate voltage. Potential opto-valleytronic and terahertz applications are discussed.
\end{abstract}

\maketitle


\section{Introduction}
In contrast to conventional semiconductors, charge carriers in graphene behave like ultra-relativistic massless particles, moving at approximately one three-hundredth of the speed of light. The relativistic nature of graphene's charge carriers leads to its highly celebrated optical and electronic properties~\cite{RevModPhys.81.109,doi:10.1126/science.1156965}, positioning graphene as a promising material for several advanced optoelectronic applications. The rise of graphene sparked a surge of interest in researching other materials whose low-energy charge carriers can be described by a Dirac-like Hamiltonian, among which are topological insulators~\cite{hasan2010colloquium,qi2011topological}, transition metal dichalcogenides~\cite{xiao2012coupled}, 3D Weyl semimetals~\cite{young2012dirac}, and materials predicted to exhibit tilted Dirac cones, such as 8--$Pmmn$ borophene~\cite{PhysRevLett.112.085502}.


Tilted Dirac cones can be classified into three categories. For a type-I cone, the equienergy contours are closed. For the other two types, the equienergy contours are open: in a type-II cone the equienergy contours are hyperbolic, while in type-III they are parabolic~\cite{10.1038_nature15768,PhysRevX.9.031010}. These various cone geometries lead to remarkably different physical properties~\cite{doi:10.1143/JPSJ.79.114715,PhysRevB.96.155418,PhysRevB.103.165415,PhysRevB.103.125425,wild2021optical,C7CP03736H,PhysRevB.97.235113,PhysRevB.97.235440,Sengupta_2018,PhysRevB.99.235413,PhysRevB.102.045417,doi:10.1143/JPSJ.77.064718,PhysRevB.91.195413,PhysRevB.96.235405,PhysRevB.98.195415,PhysRevB.99.035415,ng2021mapping,Liu_2023,PhysRevB.105.L201408,wild2023optical}. In certain materials, the Dirac cones can be naturally tilted due to the material's inherent crystal structure and properties, and there is a plethora of theoretical materials that have been predicted to possess tilted Dirac cones~\cite{PhysRevLett.112.085502,doi:10.1143/JPSJ.75.054705,PhysRevB.78.045415,PhysRevLett.105.037203,doi:10.1126/science.1256815,PhysRevX.6.041069,PhysRevB.94.195423,Ma2016,PhysRevB.95.035151,PhysRevB.95.245421,https://doi.org/10.1002/pssr.201800081,PhysRevB.98.121102,PhysRevB.100.235401,PhysRevB.100.205102,PhysRevB.102.041109}. In graphene, the tilting of Dirac cones can be induced through various means, e.g., via the application of mechanical strain
~\cite{PhysRevB.78.045415,PhysRevB.80.153412,PhysRevB.81.081407,PhysRevB.90.075406,AMORIM20161}, doping~\cite{Wang2023}, or the application of external electric~\cite{PhysRevB.77.115446,PhysRevB.81.075438,PhysRevLett.103.046808,10.1140/epjb/e2016-70605-5,PhysRevLett.103.046809,PhysRevB.79.115427,PhysRevLett.101.126804} or magnetic potentials~\cite{SOMROOB2021114501,condmat8010028}.
An alternative approach to studying systems that support tilted Dirac cones involves exploring the connection between gravitational physics and condensed-matter systems~\cite{volovik2016black,volovik2017lifshitz,hashimoto2020escape,de2021artificial,volovik2021type,konye2022horizon,konye2023anisotropic}. This approach was first considered in (3+1)-dimensional Weyl semimetals~\cite{volovik2016black}, opening the door to the investigation of black holes and gravitational lensing in a solid-state setting. In many systems with tilted Dirac cones, the spatial dependence of the tilt parameter can be achieved through various means, such as externally applied electric fields~\cite{PhysRevB.99.235150}, magnetic/pseudo-magnetic fields~\cite{farajollahpour2020synthetic, jafari2023theory}, chemical substitution~\cite{yekta2023tunning} and, in the case of a tunable circuit realization of tilted Dirac cone systems, by spatially varying the capacitor and inductor sizes~\cite{motavassal2021circuit}. The presence of a space-dependent tilt parameter in these systems allows for the exploration of nontrivial spacetime geometry and gravitomagnetic effects.
The rising interest in tilted Dirac materials has led to the revisiting of several well-known problems in graphene, such as Klein tunneling~\cite{nguyen2018klein}, atomic collapse~\cite{fu2021coulomb}, and optical absorption \cite{wild2021optical}, to now incorporate a tilt term. Transport across quasi-1D heterostructures~\cite{zhang2018oblique, zhou2019valley, yesilyurt2019electrically, zhou2020valley} in tilted Dirac materials has also been investigated, as has the alternative geometry: propagating fully confined modes along electrostatic potentials~\cite{ng2021mapping}. 

In graphene, bipolar electrostatic potentials have been demonstrated to introduce valence-like and conduction-like bands into the energy spectrum, and the size of the pseudogaps can be externally tuned through varying the top-gate voltage~\cite{PhysRevB.102.155421}. The pseudogaps are a consequence of avoided crossings, which occur due to the overlap of the spectra of the individual well and barrier that form the bipolar waveguide. In this paper, we demonstrate that bipolar waveguides in tilted Dirac materials yield spectra that significantly differ from those of materials with untilted Dirac cones. Notably, in graphene, for a given valley, the spectrum is symmetric with respect to momentum as measured about an individual K-point~\cite{PhysRevB.102.155421}. In stark contrast, when considering a given valley in a tilted Dirac material, the avoided crossings occur at different energies depending on the sign of the momentum, provided the waveguide is not normal to the tilt direction. This key difference has the potential to be harnessed for novel valleytronic applications, where the valley quantum number could serve as a basis for carrying information, akin to the role of spin in semiconductor spintronics \cite{vzutic2004spintronics,vitale2018valleytronics}.


The paper is organized as follows: first we show that the differential equations governing guided modes within a tilted Dirac waveguide can be transformed into the biconfluent Heun equation for a wide range of  potentials. We then investigate a specific case: the bound state solutions propagating along a bipolar potential formed by a well and barrier described by the Morse potential~\cite{PhysRev.34.57}. The eigenvalue spectrum of the waveguide is then determined using a simple transcendental equation. Finally, we discuss possible applications.

\

\section{The tilted Dirac equation}
The Hamiltonian describing the guided modes contained within a smooth electron waveguide in a tilted Dirac material can be written as 
\begin{equation}
\hat{H}=\hbar\left(v_{x}\sigma_{x}\hat{k}_{x}+sv_{y}\sigma_{y}\hat{k}_{y}+sv_{t}\sigma_{0}\hat{k}_{y}\right)+\sigma_{o}U(x,y)
\label{eq:Dirac_tilt_Ham}
\end{equation}
where $\hat{k}_{x}=-i\partial_{x}$, $\hat{k}_{y}=-i\partial_{y}$, $\sigma_{x,y}$ are the Pauli matrices, $\sigma_0$ is the identity matrix, $v_x$ and $v_y$ are the anisotropic velocities, $v_t$ is the tilt velocity, $s=\pm1$ is the valley index number, and $U(x,y)$ is the externally applied electrostatic potential. This Hamiltonian is of the same form as the low-energy two-band effective Hamiltonian used to describe 8--\textit{Pmmn} borophene~\cite{PhysRevB.94.165403}, 2B:\textit{Pmmn}  borophane~\cite{PhysRevB.97.125424}, and $\alpha$-(BEDT-TTF)$_{2}{I}_{3}$~\cite{doi:10.1143/JPSJ.75.054705,PhysRevB.78.045415}. In practical experiments, it is not yet possible to select the exact angle between the top gate and the crystallographic orientation. Although we will consider the specific case of aligning the waveguide along the $y$-direction, it should be emphasized that one can derive spectra for any waveguide orientation by applying the mapping method outlined in Ref.~\cite{ng2021mapping}.

When the Hamiltonian, Eq.~(\ref{eq:Dirac_tilt_Ham}), is applied to a two-component Dirac wavefunction
of the following form: $\Psi=\left(\psi_{a}\left(x\right),\,\psi_{b}\left(x\right)\right)^{\mathrm{T}}e^{ik_{y}y}$, where $\psi_{a}$ and $\psi_{b}$ are the wavefunctions associated respectively with the $a$ and $b$ sublattices of the tilted Dirac material, and
the free motion in the $y$-direction is characterized by the wave vector $k_y$ measured with respect to the Dirac point, the following coupled first-order differential equations are obtained:
\begin{equation}
\left(V-\widetilde{\varepsilon} +st\widetilde{k}_{y}\right)\psi_{a}+\left(-i\frac{\partial}{\partial z}-siT\widetilde{k}_{y}\right)\psi_{b}=0,
\label{eq:system_1}
\end{equation}
and
\begin{equation}
\left(-i\frac{\partial}{\partial z}+siT\widetilde{k}_{y}\right)\psi_{a}+\left(V-\widetilde{\varepsilon}+st\widetilde{k}_{y}\right)\psi_{b}=0,
\label{eq:system_2}
\end{equation}
where $z=x/l$, $\widetilde{k}_{y}=lk_{y}$, and $l$ is a constant, and the dimensionless tilt and anisotropic velocities are defined as $t=v_{t} / v_{x}$ and $T=v_{y}/v_{x}$, respectively. The potential, $V(z)$, and the charge carrier energy, $\widetilde{\varepsilon}$, have been scaled such that $V(z)=lU(z)/\hbar v_{x}$ and $\widetilde{\varepsilon}=l\varepsilon/\hbar v_{x}$, where $\varepsilon$ is the eigenvalue of the system. Substituting $\Psi_{+}=\psi_{a}+\psi_{b}$ and $\Psi_{-}=\psi_{a}-\psi_{b}$ into Eq.~(\ref{eq:system_1}) and Eq.~(\ref{eq:system_2}) allows them to be reduced to a  single second-order differential equation in $\Psi_{\pm}$:
\begin{equation}
    \left[\left(V-E\right)^{2}-\Delta^{2}\pm i\frac{\partial V}{\partial z}\right]\Psi_{\pm}+\frac{\partial^{2}\Psi_{\pm}}{\partial z^{2}}=0,
\label{eq:effective_SE}
\end{equation}
where $E=\widetilde{\varepsilon}-st\widetilde{k}_{y}$ and $\Delta=sT\widetilde{k}_{y}$, and the other components are found via the relation:
\begin{equation}
\Psi_{\mp}=\pm i\left(V-E\mp i\partial_{z}\right)\Psi_{\pm}/\Delta.
\label{eq:wav_second_sol}
\end{equation}
It should be noted that whenever the Schr\"{o}dinger equation possesses exact solutions for a given potential, $W(\widetilde{z})$, there exists a corresponding potential, $V(\widetilde{z})$, for which Eq.~(\ref{eq:effective_SE}) is exactly solvable, satisfying the relation $W(\widetilde{z})=(V-E)^2\pm \frac{\partial V}{\partial{\widetilde{z}}}$, where $\widetilde{z}=iz$ ~\cite{cooper1988supersymmetry}. This method can be used to obtain a broad set of zero-energy and energy-dependent potentials~\cite{ho2014zero,GHOSH2016567,schulze2017bound,Schulze-Halberg_2017}.

\section{The bipolar Morse potential}
Based upon state-of-the-art waveguides achievable in experiments~\cite{PhysRevLett.123.216804}, our proposed bipolar waveguide is created by two carbon nanotube top gates. These gates are separated by a finite distance and positioned a certain height above a metallic substrate (see Figure 1). The bipolar potential in the plane of the tilted Dirac material~\cite{PhysRevB.102.155421}, can be approximated by the piecewise function 
\begin{equation}
U(x)=\left\{ \begin{array}{cc}
-U_{0}e^{-x/l}\left(1-e^{-x/l}\right), & x>0\\
U_{0}e^{x/l}\left(1-e^{x/l}\right), & x<0
\end{array}\right.
\label{eq:model_pot}
\end{equation}
where $l$ and $U_0$ are the characteristic potential width and depth respectively.  This function is zero at the origin, possesses continuous derivatives to the first order, asymptotically approaches zero at $x=\pm\infty$, has a minimum value of $-U_0/4$  at $x= l\ln\left(2\right)$, and has a maximum value of $U_0/4$ at $x=-l\ln\left(2\right)$.


\begin{figure}[ht]
\centering
\includegraphics[width=0.8\textwidth]{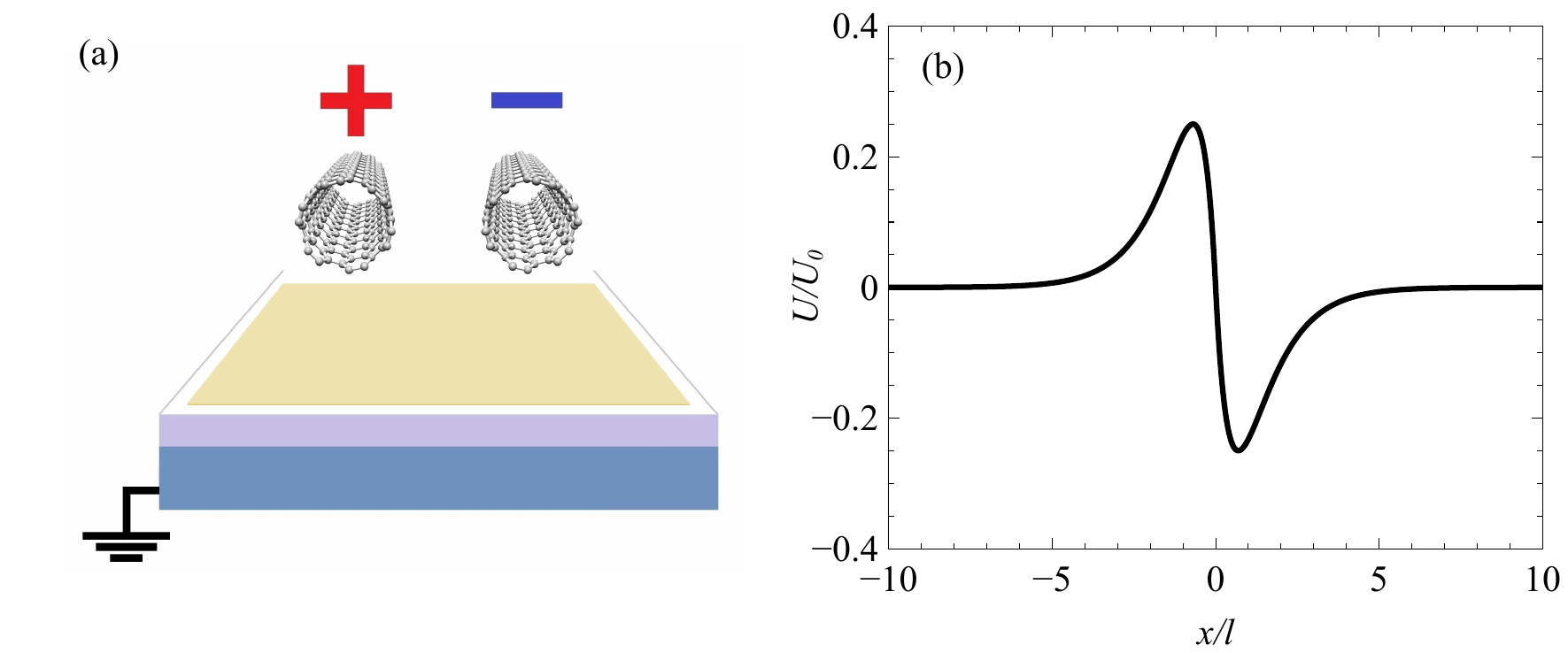}
\caption{(a) Illustration of the proposed experimental setup and (b) the bipolar potential created by a well and barrier, as described by the Morse potential: $U\left(x\right)=-\mathrm{sgn}\left(x\right)U_{0}e^{-\left|x\right|/l}\left(1-e^{-\left|x\right|/l}\right)$.
}
    \label{fig:schematic}
\end{figure}

\section{Transforming the tilted Dirac equation into the biconfluent Heun equation}
Heun's equation is a second-order linear ordinary differential equation characterized by four regular singular points~\cite{ronveaux1995heun}. Confluent forms of Heun’s differential equation occur when two or more of the regular singularities combine to form an irregular singularity. In the case of the biconfluent Heun equation, the merging of its two finite regular singular points with infinity results in one regular singularity at the origin and one irregular singularity at infinity. The biconfluent Heun equation frequently arises in various areas of mathematics and physics~\cite{ronveaux1995heun,kandemir2005two,caruso2014solving,vieira2015quantum,shahverdyan2015analytic,karwowski2014biconfluent,batic2013potentials,arda2017thermodynamic} and has been employed in numerous single-particle quantum confinement problems~\cite{ronveaux1995heun,ishkhanyan2016conditionally,ishkhanyan2016discretization,leaute1986schrodinger}. 

The normal form of the biconfluent Heun equation is given by:
\begin{equation}
\xi^{2}\frac{\partial^{2}u}{\partial\xi^{2}}+\left[\frac{1}{4}\left(1-\alpha^{2}\right)-\frac{\delta}{2}\xi+\gamma\xi^{2}-\frac{1}{4}\beta^{2}\xi^{2}-\beta\xi^{3}-\xi^{4}\right]u=0.
\label{eq:norm_biheun}
\end{equation}
This equation has the solution 
$u\left(\xi\right)=\xi^{\frac{1}{2}\left(1+\alpha\right)}e^{-\frac{1}{2}\xi\left(\beta+\xi\right)}H\left(\xi\right)$, where $H(\xi)=H(\alpha, \beta, \gamma, \delta; \xi)$ are the solutions to the canonical biconfluent Heun equation~\cite{ronveaux1995heun}. In what follows we shall perform transformations on the dependent and independent variables of Eq.~(\ref{eq:effective_SE}), and find the corresponding energy-independent potentials which allow Eq.~(\ref{eq:effective_SE}) to be converted into the normal form of the biconfluent Heun equation. A similar approach has been implemented to determine the bound states of charge carriers in two-dimensional Dirac materials for a family of one-dimensional confining potentials~\cite{PhysRevA.102.052229} using the confluent Heun equation. It should also be noted that there exists a broad range of potentials for which the Schr\"{o}dinger equation reduces to the Heun equation or one of its confluent  forms~\cite{ishkhanyan2015thirty,ishkhanyan2016schrodinger,ishkhanyan2016discretization,ishkhanyan2018schrod}.



Applying the change of variable $\xi=\xi\left(z\right)$, satisfying $\frac{\partial\xi}{\partial z}=-c_2\xi^{n}$, where $c_2$ is a constant, and selecting a wave function of the form $\Psi_{\pm}=\psi_{\pm}\xi^{-\frac{n}{2}}$, allows Eq.~(\ref{eq:effective_SE}) to be expressed as
\begin{equation}
    \frac{\partial^{2}\psi_{\pm}}{\partial\xi^{2}}+\left[\frac{\left(V-E\right)^{2}-\Delta^{2}\mp ic_2\xi^{n}\frac{\partial V}{\partial\xi}}{c_2^{2}\xi^{2n}}+\frac{\frac{1}{2}n\left(1-\frac{1}{2}n\right)}{\xi^{2}}\right]\psi_{\pm}=0.
    \label{eq:tobereduced}
\end{equation}
Let us consider the case of $n=1$, i.e., $\xi=c_{1}e^{-c_2z}$. In this instance if the potential is of the form 
\begin{equation}
V=a_{2}\xi^{2}+a_{1}\xi^{1}
\label{eq:pot_tobe}
\end{equation}
where $a_2=s_{c}ic_{2}$ with $s_c$ taking the value of $1$ or $-1$, then Eq.~(\ref{eq:tobereduced}) reduces to the normal form of the biconfluent Heun equation, Eq.~(\ref{eq:norm_biheun}), with
\begin{equation*}
\alpha=s_{\alpha}\frac{2\sqrt{\Delta^{2}-E^{2}}}{c_{2}},\qquad\beta=-s_{c}\frac{2a_{1}}{c_{2}}i,
\end{equation*}
\begin{equation*}
\gamma=2s_{c}\frac{\left(\pm c_{2}-iE\right)}{c_{2}},\qquad\delta=\frac{2a_{1}\left(2E\pm ic_{2}\right)}{c_{2}^{2}},
\end{equation*}
where $s_\alpha$ can take the value of $1$ or $-1$. When $a_{1}=-s_{c}c_{1}c_{2}i$ and $V_{0}=s_{c}c_{1}^{2}i$ then the potential takes the desired form of
\begin{equation}
V=-c_{2}V_{0}e^{-c_{2}z}\left(1-e^{-c_{2}z}\right),
\end{equation}
where $c_2$ is now restricted such that $c_2=1$ for $z>0$, and $c_2=-1$ for $z<0$, 
and in this instance the linearly independent solutions to Eq.~(\ref{eq:effective_SE}) are
\begin{equation}
\Psi_{\pm, s_c}=\sum_{s_{\alpha}}A_{\pm,\,s_{\alpha},\,s_{c}}\xi^{\frac{1}{2}\alpha}e^{-\frac{1}{2}\xi\left(\beta+\xi\right)}H\left(\alpha,\,\beta,\,\gamma,\,\delta;\,\xi\right),
\label{eq:wavefunctions}
\end{equation}
where $A_{\pm,\,s_{\alpha},\,s_{c}}$ is a constant, $\alpha=s_{\alpha}c_{2}2\sqrt{\Delta^{2}-E^{2}}$, $\beta=-2c_{1}$, $\gamma=2\left(\pm1 -ic_{2}E\right)$, $\delta=2c_{1}\left(\pm 1 -i2c_{2}E\right)$, $c_1=\sqrt{-iV_{0}}$, and $V_0$ is a positive parameter. 

There are several other cases where the tilted Dirac equation, Eq.~(\ref{eq:Dirac_tilt_Ham}), can be reduced to the normal form of the biconfluent Heun equation. For example, when $n=-1$, the energy-independent potential formed by a linear combination of an inverse-square-root and Coulomb potential
\begin{equation}
V\left(x\right)=\frac{V_{1}}{\sqrt{x-x_{0}}}+\frac{V_{2}}{x-x_{0}},
\end{equation}
where $x_0$, $V_1$, and $V_2$ are constants, can be solved in terms of biconfluent Heun functions. It should be noted that both the shifted Coulomb and inverse-square-root potentials were previously solved for the one-dimensional Dirac problem in terms of Whittaker and Hermite functions, respectively~\cite{PhysRevA.90.052116,ishkhanyan2020exact,ishkhanyan2023exact}. 
Another example is for the case of $n=0$; in this instance, the potential formed by the addition of a Coulomb and linear potential
\begin{equation}
V\left(x\right)=\frac{V_{1}}{x}+V_{2}x,
\end{equation}
where $V_1$, and $V_2$ are energy-independent constants, can be solved in terms of biconfluent Heun functions. For the case of $V_1=0$, Eq.~(\ref{eq:Dirac_tilt_Ham}) can also be solved in terms of Hermite functions~\cite{dominguez1990solvable,chao2007exact}.


\section{Guided modes in a bipolar Morse potential}
We shall now analyze the asymptotic behavior of the wavefunctions as $\left|z\right|$ approaches infinity, i.e., $\xi\rightarrow0$. It can be seen from Eq.~(\ref{eq:wavefunctions}) that the asymptotic expressions for $\Psi_{\pm}$ become 
$\Psi_{\pm}=\sum_{s_{\alpha}}A_{\pm,\,s_{\alpha},\,s_{c}}\xi^{\frac{1}{2}\alpha}$. Thus, for a bound state to exist, it is necessary that $\alpha$ is real and positive (i.e., $\Delta^2>E^2$), implying that $A_{\pm,\,s_{\alpha},\,s_{c}}=0$ unless $s_\alpha=s_c$. The relationship between the remaining nonzero coefficients of $A_{\pm,\,s_{\alpha},\,s_{c}}$ in the well-like and barrier-like regions is determined by the boundary conditions at the origin, where we require both continuity and smoothness of the wavefunction:
\begin{equation}
\Psi_{-1}\left(z=0\right)=\Psi_{1}\left(z=0\right),\qquad\left.\frac{\partial\Psi_{-1}}{\partial z}\right|_{z=0}=\left.\frac{\partial\Psi_{1}}{\partial z}\right|_{z=0}.
\label{eq:boundary_cond}
\end{equation}
For brevity, we have omitted the $\pm$ sign appearing in Eq.~(\ref{eq:wavefunctions}), leaving only the index $c_2$, where $c_2=-1$ for $z<0$ and $c_2=1$ for $z>0$. Therefore, the eigenvalues are determined via the zeros of the Wronskian, $W\left(\Psi_{1},\,\Psi_{-1}\right)(z=0)$. Notably, the Wronskian method has been successfully employed to determine the energy spectrum for guided modes in several Dirac waveguides~\cite{PhysRevA.102.052229,PhysRevA.89.012101,10.1038/s41598-017-11411-w} and eigenstates for the quantum Rabi model~\cite{xie2017quantum,zhong2013analytical,maciejewski2014full}.

\begin{figure}
    \centering
    \includegraphics[width=0.5\textwidth]{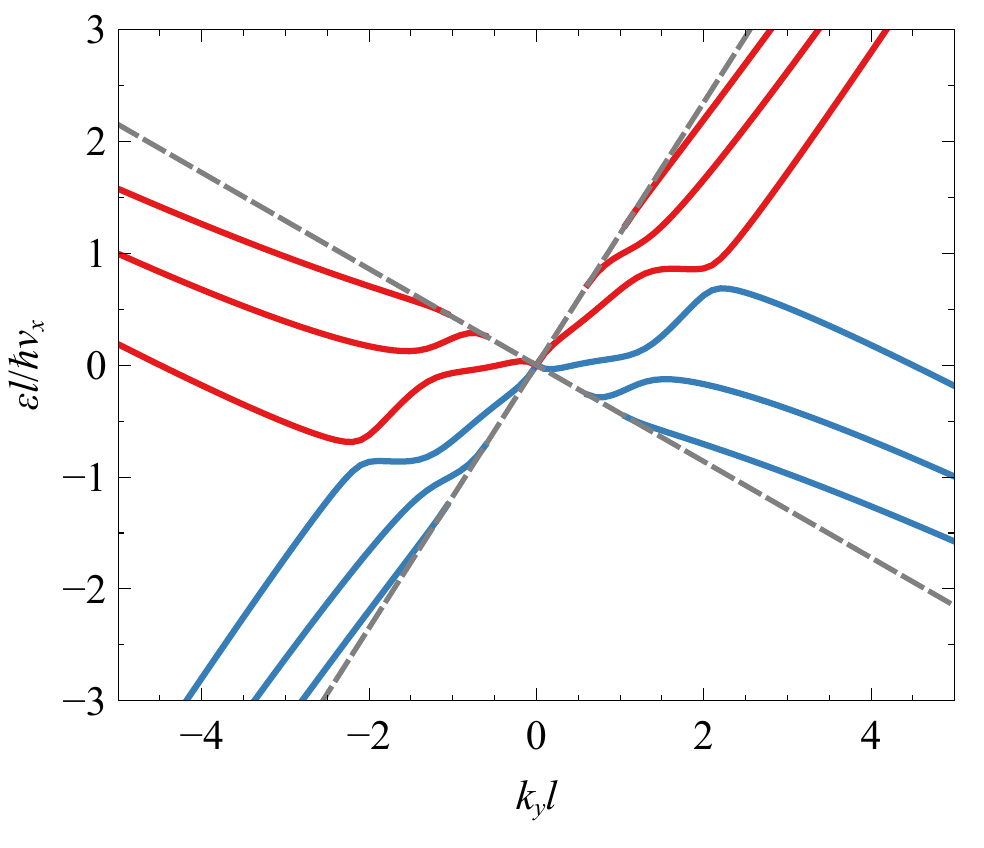}
    \caption{
The energy spectrum of guided modes (for the $s=1$ valley) in a bipolar Morse potential, $U\left(x\right)=-\mathrm{sgn}\left(x\right)U_{0}e^{-\left|x\right|/l}\left(1-e^{-\left|x\right|/l}\right)$, of strength $V_0= l U_{0} / \hbar v_x= 10$, as a function of wavenumber along the waveguide in a tilted Dirac material, defined by parameters $v_x=0.86\,v_{\mathrm{F}}$, $v_y=0.69\,v_{\mathrm{F}}$ and $v_t=0.32\,v_{\mathrm{F}}$, where $v_{\mathrm{F}}$ is the Fermi velocity of graphene. Here, only the three lowest modes associated with the individual well (red lines) and the three highest modes associated with the individual barrier (blue lines) that form the bipolar waveguide are displayed. The grey dashed lines indicate the boundary at which the bound states merge with the continuum.
}
\label{fig:spectra_1}
\end{figure}

In Figure~\ref{fig:spectra_1} we plot the eigenvalues obtained from the zeros of the Wronskian $W\left(\Psi_{1},\,\Psi_{-1}\right)(z=0)$, for a bipolar potential with characteristic potential strength $V_0=10$, for a tilted Dirac material (8-\textit{Pmmn} borophene) defined by parameters $v_x=0.86\,v_{\mathrm{F}}$, $v_y=0.69\,v_{\mathrm{F}}$
and $v_t=0.32\,v_{\mathrm{F}}$, where $v_{\mathrm{F}}$ is the Fermi velocity of graphene. The boundary at which the bound states merge with the continuum is denoted by the grey dashed lines. In stark contrast to graphene, the presence of the tilt term breaks the $\varepsilon(k_y) = \varepsilon(-k_y)$ symmetry for a given valley. As a consequence, avoided crossings occur at different energies for opposite signs of $k_y$. This asymmetry in the band structure opens up opportunities for potential valleytronic applications. For example, if the Fermi level is positioned within the higher-energy pseudogap, any charge carriers optically excited across the gap will result in carriers propagating along a specific direction of the waveguide that are fully valley-polarized. This phenomenon does not occur in graphene bipolar waveguides, as pseudogaps occur at the same energy.

\begin{figure}[h]
\centering
\includegraphics[width=0.75\textwidth]{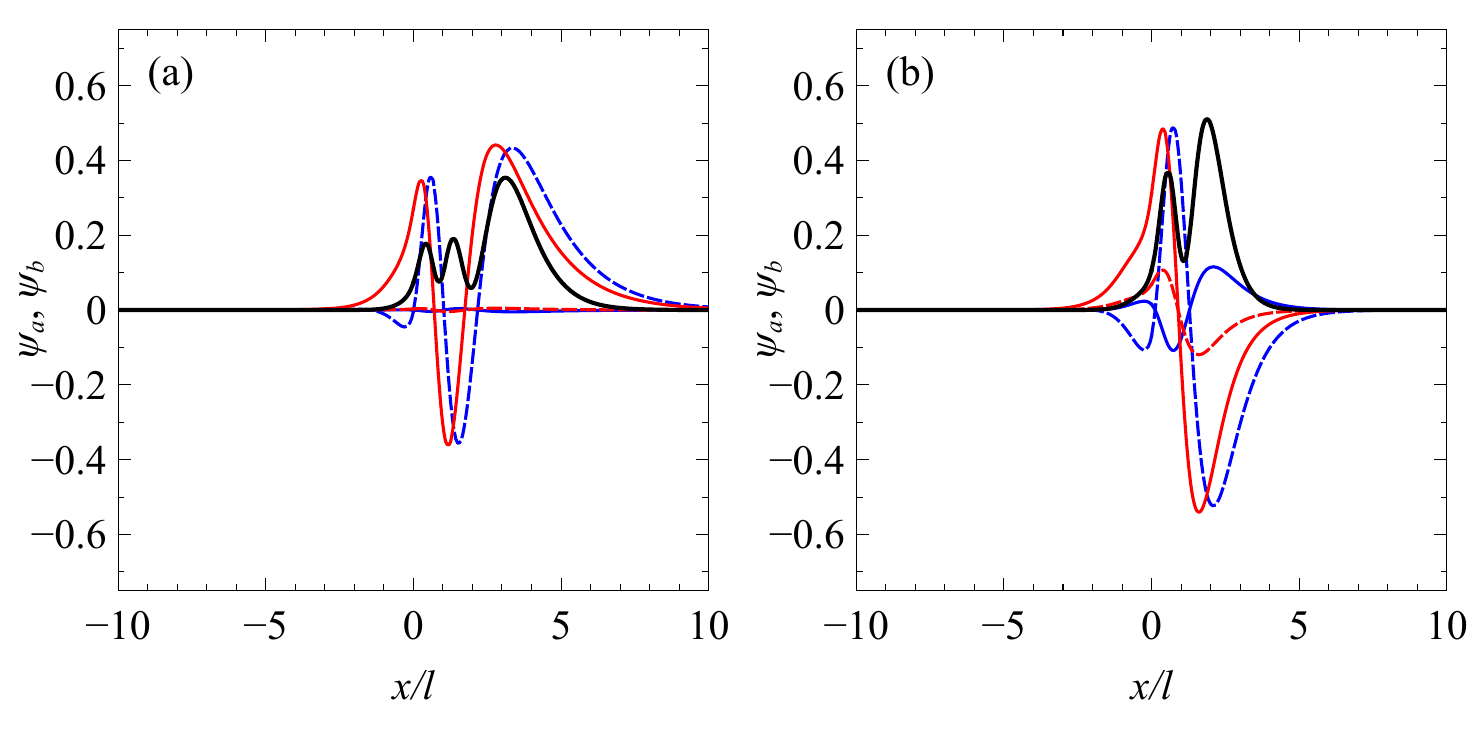}
\includegraphics[width=0.75\textwidth]{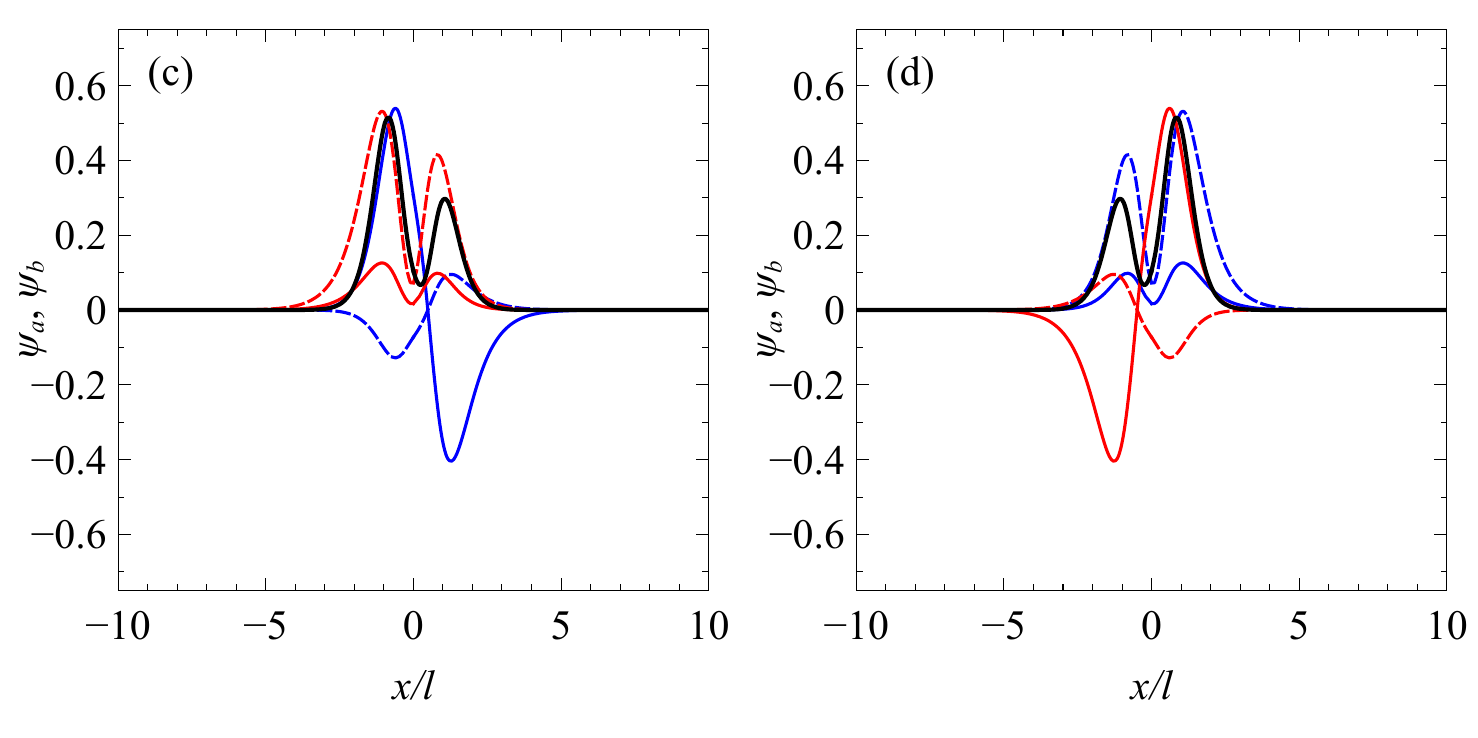}
\includegraphics[width=0.75\textwidth]{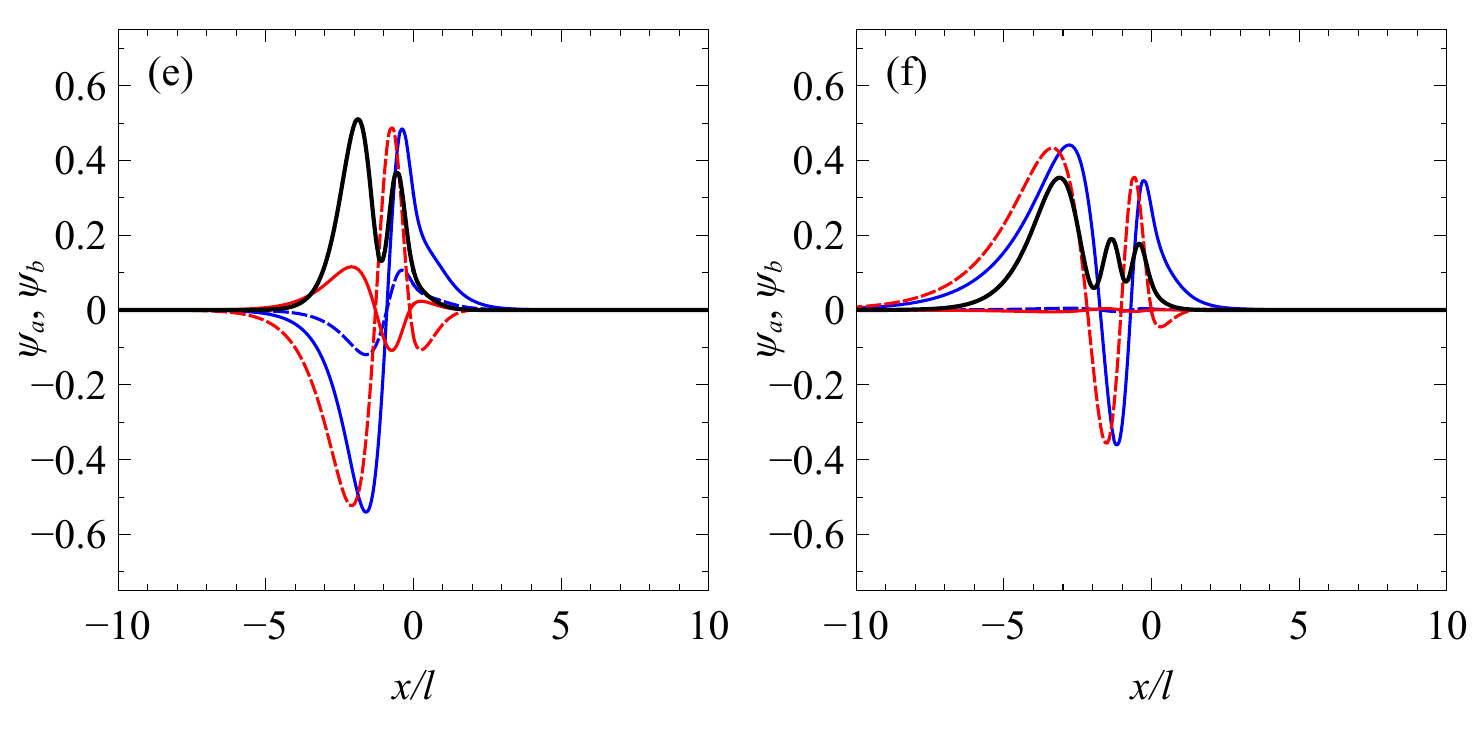}
\caption{
The normalized wavefunctions of guided modes (for the $s=1$ valley) in a bipolar Morse potential, $U\left(x\right)=-\mathrm{sgn}\left(x\right)U_{0}e^{-\left|x\right|/l}\left(1-e^{-\left|x\right|/l}\right)$, of strength $V_0= l U_{0} / \hbar v_x= 10$, for $k_yl=2$ and energy: (a) $\widetilde{\varepsilon}=2.195$, (b) $\widetilde{\varepsilon}=1.658$, (c) $\widetilde{\varepsilon}=0.864$, (d) $\widetilde{\varepsilon}=0.624$, (e) $\widetilde{\varepsilon}=-0.169$, and (f) $\widetilde{\varepsilon}=-0.706$, in a tilted Dirac material defined by parameters $v_x=0.86\,v_{\mathrm{F}}$, $v_y=0.69\,v_{\mathrm{F}}$ and $v_t=0.32\,v_{\mathrm{F}}$. 
The solid red and blue lines correspond to real parts of $\psi_a$ and $\psi_b$, respectively, while the dashed lines correspond to their imaginary parts. The black line shows the electron density $\left|\Psi\right|^{2}=\left|\psi_{a}\right|^{2}+\left|\psi_{b}\right|^{2}$.
}
\label{fig:wave_plots}
\end{figure}

It should be noted that, if the bipolar potential is an odd function, then it can be seen from Eqs.~(\ref{eq:system_1}) and~(\ref{eq:system_2}) that for a given $k_y$, the eigenfunctions of $E$ and $-E$ are related through the simple relations $\psi_{a}\left(-E,z\right)=\psi_{b}\left(E,-z\right)$ and $\psi_{b}\left(-E,z\right)=\psi_{a}\left(E,-z\right)$. Therefore, the electron density of the modes associated with the individual well (eigenfunctions of $E$) and barrier (eigenfunctions of $-E$) which form the bipolar waveguide are mirror images of each other (see Figure~\ref{fig:wave_plots}). Thus, away from the pseudogap, the positive $E$ eigenfunctions have an electron-like character (i.e., mainly localized in the proximity of the well), and the negative $E$ eigenfunctions have a hole-like character (i.e., mainly localized in the proximity of the barrier). Another characteristic of an odd bipolar potential is that, for a given $E$, the wavefunctions of the other valley can be obtained by a simple change of sign in $k_y$, along with the switching of the spinor components $\psi_a$ and $\psi_b$. In other words: $\left(\psi_{a}\left(E,-k_{y}\right),\psi_{b}\left(E,-k_{y}\right)\right)^{\mathrm{T}}=\left(\psi_{b}\left(E,k_{y}\right),\psi_{a}\left(E,k_{y}\right)\right)^{\mathrm{T}}$. Furthermore, if a bound-state solution exists for a given $E$ and $\Delta$, then a solution also exists for $E$ and $-\Delta$. Therefore, the eigenvalue spectrum of the $s=-1$ valley can be obtained by reflecting the eigenvalue spectrum of the $s=1$ valley about $k_y=0$~\cite{ng2021mapping}.

Although the exact optical selection rules of a tilted Dirac material depend on the details of the lattice structure, for materials where optical transitions across the pseudogap are allowed, the presence of van Hove singularities at the pseudogap edges should lead to a very sharp absorption maximum. This, combined with the ability to control the size of the pseudogap via the applied top-gate voltage, gives rise to the possibility of gate-controlled polarization-sensitive THz detectors with a very high sensitivity of the photocurrent to photon frequency.

\section{Conclusions}
We have shown that the (2+1)-dimensional massless Dirac equation containing a tilt term can be reduced to the biconfluent Heun equation for a range of energy-independent electrostatic potentials. We studied the particular case of a bipolar potential formed by a well and barrier, modeled by the Morse potential. Our study shows that the ability of a tilted Dirac material to localize charge carriers in both a barrier and a well allows for the creation of pseudogaps in the material's spectrum. The size of these pseudogaps can be controlled by the strength of the guiding potentials. In stark contrast to graphene, the presence of a tilt term leads to the breaking of reflection symmetry, causing them to occur at different energies within a given valley. These findings have direct relevance to the expanding field of research on tilted Dirac materials, as well as applications in opto-valleytronics and terahertz science and technologies.

\ack
This work was supported by the EU H2020-MSCA-RISE projects TERASSE (Project No. 823878) and CHARTIST (Project No. 101007896) as well as by the NATO Science for Peace and Security project NATO.SPS.MYP.G5860.  R. R. H. acknowledges financial support from URCO (Grant No. 15 F 2TAY21–3TAY22). M. E. P. acknowledges support from UK EPSRC (Grant No. EP/Y021339/1).


\section*{Data availability statement}
No new data were created or analysed in this study.

\section*{Ethical approval}
The subject of this study is mathematical physics and complies with the journal’s ethical rules.

\section*{Competing interests}
I declare that the authors have no competing interests as defined by IOP, or other interests that might be perceived to influence the results and/or discussion reported in this paper

\section*{Bibliography}
\bibliographystyle{iopart-num}
\bibliography{references}
\end{document}